\documentclass[twoside,twocolumn,9pt]{article}
\usepackage{extsizes}
\usepackage[super,sort&compress,comma]{natbib} 
\usepackage[version=3]{mhchem}
\usepackage[left=1.5cm, right=1.5cm, top=1.785cm, bottom=2.0cm]{geometry}
\usepackage{balance}
\usepackage{mathptmx}
\usepackage{sectsty}
\usepackage{graphicx} 
\usepackage{lastpage}
\usepackage[format=plain,justification=justified,singlelinecheck=false,font={stretch=1.125,small,sf},labelfont=bf,labelsep=space]{caption}
\usepackage{float}
\usepackage{fancyhdr}
\usepackage{fnpos}
\usepackage[english]{babel}
\addto{\captionsenglish}{%
  \renewcommand{\refname}{Notes and references}
}
\usepackage{array}
\usepackage{droidsans}
\usepackage{charter}
\usepackage[T1]{fontenc}
\usepackage[usenames,dvipsnames]{xcolor}
\usepackage{setspace}
\usepackage[compact]{titlesec}
\usepackage{hyperref}
\usepackage{epstopdf}%This line makes .eps figures into .pdf - please comment out if not required.

\usepackage{xcolor}
\usepackage{soul}

\definecolor{cream}{RGB}{222,217,201}

\begin{document}

\pagestyle{fancy}
\thispagestyle{plain}
\fancypagestyle{plain}{
%%%HEADER%%%
\renewcommand{\headrulewidth}{0pt}
}
%%%END OF HEADER%%%

%%%PAGE SETUP - Please do not change any commands within this section%%%
\makeFNbottom
\makeatletter
\renewcommand\LARGE{\@setfontsize\LARGE{15pt}{17}}
\renewcommand\Large{\@setfontsize\Large{12pt}{14}}
\renewcommand\large{\@setfontsize\large{10pt}{12}}
\renewcommand\footnotesize{\@setfontsize\footnotesize{7pt}{10}}
\makeatother

\renewcommand{\thefootnote}{\fnsymbol{footnote}}
\renewcommand\footnoterule{\vspace*{1pt}% 
\color{cream}\hrule width 3.5in height 0.4pt \color{black}\vspace*{5pt}} 
\setcounter{secnumdepth}{5}

\makeatletter 
\renewcommand\@biblabel[1]{#1}            
\renewcommand\@makefntext[1]% 
{\noindent\makebox[0pt][r]{\@thefnmark\,}#1}
\makeatother 
\renewcommand{\figurename}{\small{Fig.}~}
\sectionfont{\sffamily\Large}
\subsectionfont{\normalsize}
\subsubsectionfont{\bf}
\setstretch{1.125} %In particular, please do not alter this line.
\setlength{\skip\footins}{0.8cm}
\setlength{\footnotesep}{0.25cm}
\setlength{\jot}{10pt}
\titlespacing*{\section}{0pt}{4pt}{4pt}
\titlespacing*{\subsection}{0pt}{15pt}{1pt}
%%%END OF PAGE SETUP%%%

%%%FOOTER%%%
\fancyfoot{}
\fancyfoot[LO,RE]{\vspace{-7.1pt}\includegraphics[height=9pt]{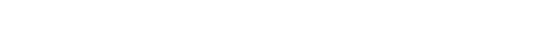}}
\fancyfoot[CO]{\vspace{-7.1pt}\hspace{11.9cm}\includegraphics{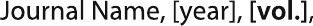}}
\fancyfoot[CE]{\vspace{-7.2pt}\hspace{-13.2cm}\includegraphics{head_foot/RF}}
\fancyfoot[RO]{\footnotesize{\sffamily{1--\pageref{LastPage} ~\textbar  \hspace{2pt}\thepage}}}
\fancyfoot[LE]{\footnotesize{\sffamily{\thepage~\textbar\hspace{4.65cm} 1--\pageref{LastPage}}}}
\fancyhead{}
\renewcommand{\headrulewidth}{0pt} 
\renewcommand{\footrulewidth}{0pt}
\setlength{\arrayrulewidth}{1pt}
\setlength{\columnsep}{6.5mm}
\setlength\bibsep{1pt}
%%%END OF FOOTER%%%

%%%FIGURE SETUP - please do not change any commands within this section%%%
\makeatletter 
\newlength{\figrulesep} 
\setlength{\figrulesep}{0.5\textfloatsep} 

\newcommand{\topfigrule}{\vspace*{-1pt}% 
\noindent{\color{cream}\rule[-\figrulesep]{\columnwidth}{1.5pt}} }

\newcommand{\botfigrule}{\vspace*{-2pt}% 
\noindent{\color{cream}\rule[\figrulesep]{\columnwidth}{1.5pt}} }

\newcommand{\dblfigrule}{\vspace*{-1pt}% 
\noindent{\color{cream}\rule[-\figrulesep]{\textwidth}{1.5pt}} }

\makeatother
%%%END OF FIGURE SETUP%%%

%%%TITLE, AUTHORS AND ABSTRACT%%%
\twocolumn[
  \begin{@twocolumnfalse}
{\includegraphics[height=30pt]{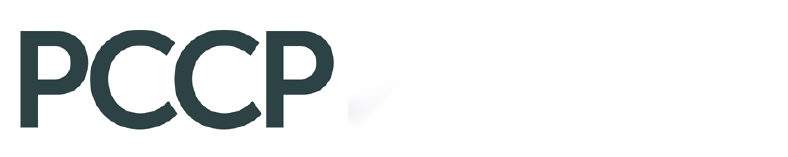}\hfill\raisebox{0pt}[0pt][0pt]{\includegraphics[height=55pt]{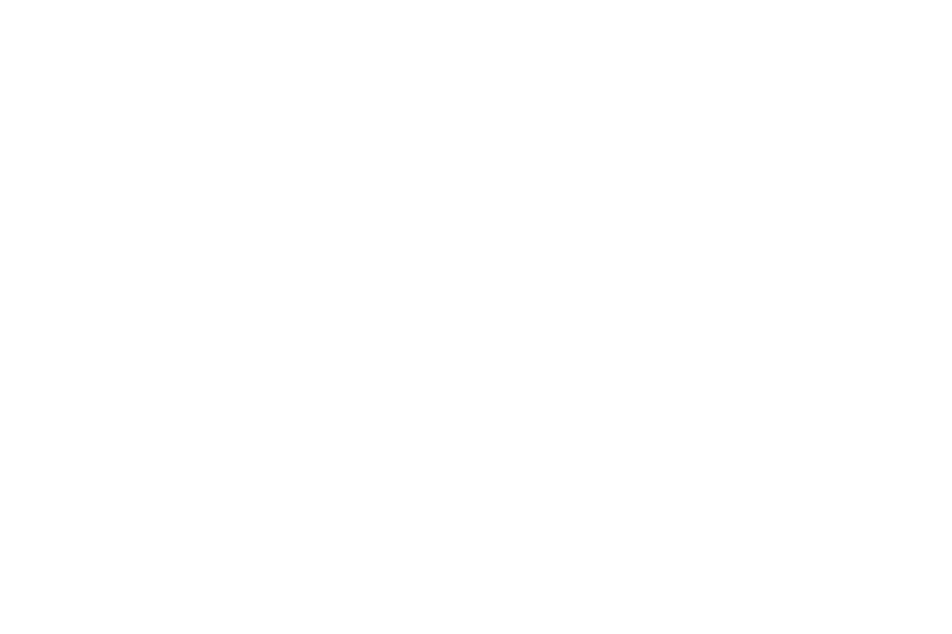}}\\[1ex]
\includegraphics[width=18.5cm]{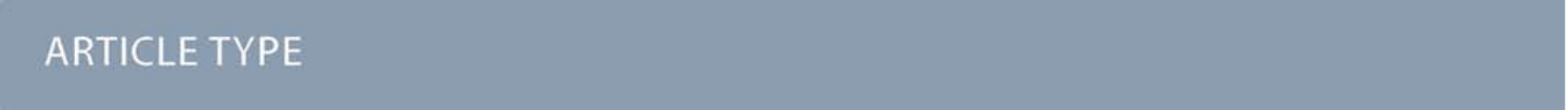}}\par
\vspace{1em}
\sffamily
\begin{tabular}{m{4.5cm} p{13.5cm} }

\includegraphics{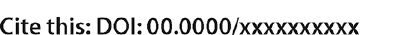} & \noindent\LARGE{\textbf{Optimizing Pulsed-Laser Ablation Production of AlCl Molecules for Laser Cooling}} \\
\vspace{0.3cm} & \vspace{0.3cm} \\

 & \noindent\large{Taylor N. Lewis,$^{a\dag}$ Chen Wang,$^{b\dag}$ John R. Daniel,$^{b}$ Madhav Dhital,$^{b}$ Christopher J. Bardeen,$^{a\ast}$ and Boerge Hemmerling$^{b\ast}$} \\

\includegraphics{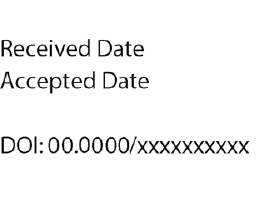} & \noindent\normalsize{Aluminum monochloride (AlCl) has been proposed as a promising candidate for laser cooling to ultracold temperatures, and recent spectroscopy results support this prediction. It is challenging to produce large numbers of AlCl molecules because it is a highly reactive open-shell molecule and must be generated in situ.  Here we show that pulsed-laser ablation of stable, non-toxic mixtures of Al with an alkali or alkaline earth chlorides, denoted XCl$_n$, can provide a robust and reliable source of cold AlCl molecules.  Both the chemical identity of XCl$_n$ and the Al:XCl$_n$ molar ratio are varied, and the yield of AlCl is monitored using absorption spectroscopy in a cryogenic gas.  For KCl, the production of Al and K atoms was also monitored.  We model the AlCl production in the limits of nonequilibrium recombination dominated by first-encounter events.  The non-equilibrium model is in agreement with the data and also reproduces the observed trend with different XCl$_n$ precursors.  We find that AlCl production is limited by the solid-state densities of Al and Cl atoms and the recondensation of Al atoms in the ablation plume.  We suggest future directions for optimizing the production of cold AlCl molecules using laser ablation.} \\

\end{tabular}

 \end{@twocolumnfalse} \vspace{0.6cm}]
%%%END OF TITLE, AUTHORS AND ABSTRACT%%%

%%%FONT SETUP - please do not change any commands within this section
\renewcommand*\rmdefault{bch}\normalfont\upshape
\rmfamily
\section*{}
\vspace{-1cm}

%%%FOOTNOTES%%%

\footnotetext{\textit{$^{a}$~Department of Chemistry, University of California, Riverside, CA 92521, USA. E-mail: christob@ucr.edu}}
\footnotetext{\textit{$^{b}$~Department of Physics and Astronomy, University of California, Riverside, CA 92521, USA. E-mail: boergeh@ucr.edu }}
\footnotetext{$\dag$~These authors contributed equally to this work.}
\footnotetext{$\ast$~Corresponding authors}

%Please use \dag to cite the ESI in the main text of the article.
%If you article does not have ESI please remove the the \dag symbol from the title and the footnotetext below.
\footnotetext{\ddag~Electronic Supplementary Information (ESI) available: Target mapping, simulated XCl$_n$ R$_{mol}$ curves, average absorption compared to models, and Al density calculation. See DOI: 10.1039/cXCP00000x/}
%additional addresses can be cited as above using the lower-case letters, c, d, e... If all authors are from the same address, no letter is required

%%%END OF FOOTNOTES%%%

%%%MAIN TEXT%%%%
\section{Introduction}
Laser cooling of atoms \cite{Wineland1978,Phillips1998,Cohen-Tannoudji1998,Chu1998,Eschner2003} has led to significant advances in fundamental physics, including the creation of Bose-Einstein condensates\cite{Davis1995,Anderson1995} and the demonstration of quantum phase transitions\cite{Greiner2002}. It also provides the basis for many precision tests of fundamental theories \cite{Uzan2003,Baker2021} and atomic clocks\cite{Ludlow2015}.
Extending laser cooling techniques to molecules could open up completely new directions of research, such as controlled chemical reactions\cite{Krems2008,Ni2010,Ospelkaus2010,Ye2018,Tscherbul2020,Liu2022}, quantum simulation of strongly interacting systems\cite{Carr2009,Micheli2006}, searches for physics beyond the Standard Model, and precision tests of fundamental theories\cite{Andreev2018,Cairncross2017,Kozyryev2017a,Hudson2011,Kozyryev2021,Kondov2019,DeMille2008,Chin2009,Kajita2009,Beloy2010,Jansen2014,Dapra2016,Kobayashi2019}.
The large electric dipole moments of polar molecules also makes possible the creation of arrays of entangled molecular qubits that have been proposed as a novel platform for quantum computing \cite{DeMille2002,Yelin2006,Yu2019}.

Given these potential applications, tremendous experimental effort has been put into the field in recent years, resulting in successful laser cooling and magneto-optical trapping of several diatomic species such as CaF\cite{Anderegg2017,Truppe2017,Williams2017}, SrF\cite{Barry2014}, and  YO\cite{Collopy2018}. Computational results and experimental studies suggest there exist other molecules with excellent properties for
laser cooling and trapping \cite{DiRosa2004}, including BaH \cite{Iwata2017}, CaOH \cite{Baum2020,Augenbraun2021}, YbF \cite{Lim2018}, CaOCH$_3$ \cite{Mitra2020}, YbOH \cite{Augenbraun2020,Kozyryev2017a}, and SrOH \cite{Kozyryev2017}, which have been laser cooled. Other proposed diatomics for laser cooling are AlF \cite{Truppe2019,Doppelbauer2021}, BaF \cite{Chen2017,Albrecht2020}, Cs$_2$ \cite{Bahns1996}, MgF \cite{Xu2016}, RaF \cite{Isaev2010}, TiO \cite{Stuhl2008}, TlF \cite{Norrgard2017}, and CH \cite{Schnaubelt2021}.
AlCl has been predicted to have excellent properties for laser cooling and trapping, including a large Franck-Condon factor of 99.88\% and strong cycling.\cite{Wan2016, Yang2016, Ren2020} High resolution spectroscopy experiments have recently confirmed these theoretical predictions, providing additional motivation to study this molecule \cite{Daniel2021}.
One prerequisite for the study of AlCl is to have a general and versatile technique to produce a large number of molecules in the gas phase, especially for experiments that aim to create quantum degenerate molecular gases.  

AlCl is an unstable molecule that must be created in situ. 
The production of gas phase AlCl for spectroscopic measurements has most commonly been accomplished by heating or putting a discharge through AlCl$_3$\cite{Bhaduri1934,Sharma1951,Tsunoda1978,Ram1982,Mahieu1989,Mahieu1989a,Futerko1993,Saksena1998}.
High temperature ovens have also been used to react Al vapor with separate sources of Cl atoms, including gases like Cl$_2$ and HCl,\cite{Futerko1993,Rogowski2002,Himmel2005} as well as vaporized solids like MgCl$_2$, AgCl, and AlCl$_3$\cite{Lide1965,Wyse1972,Hedderich1993,Dearden1993}.  
These high temperature sources produce translationally hot molecules and generate a heavy thermal load. In molecular laser cooling experiments that operate on a cryogenic buffer-gas cell, typically laser ablation is used to minimize the heat load. This approach was used to generate AlCl for the recent spectroscopy studies \cite{Daniel2021}, but there has been no systematic exploration of different conditions and precursors for optimal AlCl production. 
In optimizing the production of AlCl by laser ablation, several factors need to be taken into consideration for the solid target, including ease of preparation, safety, and the yield of gas phase AlCl. 
From previous work on thermal sources, the most obvious choice for a target would be neat AlCl$_3$, but this material presents several practical difficulties.  It rapidly decomposes in the presence of water vapor, producing toxic HCl gas and inert Al$_2$O$_3$.  The rate of this composition depends on how the AlCl$_3$ is stored, and the details of sample loading (time of air exposure, relative humidity). Although it is straightforward to press AlCl$_3$ into a pellet, we found that these chemically unstable targets provided highly variable AlCl signals from run to run.

In an effort to generate more reproducible results, we explored mixtures of Al with a stable ionic compound source of chloride, denoted XCl$_n$, where X is the cation and $\emph{n}$ is the number of associated Cl anions.  The laser ablation process involves several chemical steps, including rapid non-equilibrium heating of the solid target, volatilization of the precursors by breaking Al-Al and X-Cl bonds, diffusion, and finally Al-Cl bond formation.  
Any one of these steps could act as a bottleneck for AlCl production.  In this work, our goal is to gain a better understanding of how this process works in order to optimize AlCl production for future laser cooling experiments.  Both the chemical identity of XCl$_n$ and the Al:XCl$_n$ molar ratio are varied, and the yield of AlCl is monitored using absorption spectroscopy in a cryogenic buffer-gas beam  cell.\cite{Hutzler2012}  The production of Al and K atoms was also monitored for Al:KCl mixtures.  

We develop a simple framework to describe AlCl production in the limit of  nonequilibrium reaction dynamics dominated by first-encounter events.  With the additional assumption that Al atom production is partially suppressed by recombination under the cryogenic conditions, this model provides a quantitative description of the data and reproduces the observed trend with different XCl$_n$ precursors.  The general conclusion is that using Al:XCl$_n$ mixtures as ablation targets provides a robust and general strategy for AlCl production.  This preliminary investigation of AlCl formation in ablation plumes should help provide a guide for the development of new ablation precursors and their optimization.  
The ultimate goal is to develop convenient, high efficiency sources of metastable diatomic molecules that are amenable to cooling to nano-Kelvin temperatures. 

\section{Experimental}
\subsection{Target Preparation}
Aluminum powder and various chloride sources were mixed to make pellets for target analysis. To analyze the molar mixing ratios, 99\% BioXtra potassium chloride (KCl, Sigma Aldrich) was mixed with 99.95\% aluminum powder $< 75\,\mu$m (Sigma Aldrich) in increasing Al:Cl molar ratios from 1:25 to 10:1. A 1:4 molar ratio of Al:Cl was then used for the other chloride sources, crystalline sodium chloride (NaCl, Fisher Scientific), 98\% anhydrous magnesium chloride (MgCl$_2$, Sigma Aldrich), and anhydrous calcium chloride (CaCl$_2$, Fisher Scientific). For the AlCl$_3$ sample, pure 98\% sublimed, anhydrous aluminum trichloride (Sigma Aldrich) was used. Each of the powder mixtures were put into a 12\,mm pellet die between two anvils and pelletized using a hydraulics press with 6000 psi for 1 minute. A thin layer of Stycast 2850FT epoxy was used to glue the sample pellets onto a copper target holder.  For the Al:XCl$_n$ mixtures, the sample was allowed to cure for 24 hours in air, while the AlCl$_3$ samples were wrapped in parafilm and allowed to cure for 4-5 hours.  After the epoxy hardened, the target holder was loaded into the copper cell at the heart of the ablation chamber. 

\subsection{Spectroscopy}
\begin{figure}[h]
\centering
  \includegraphics[height=3.5cm]{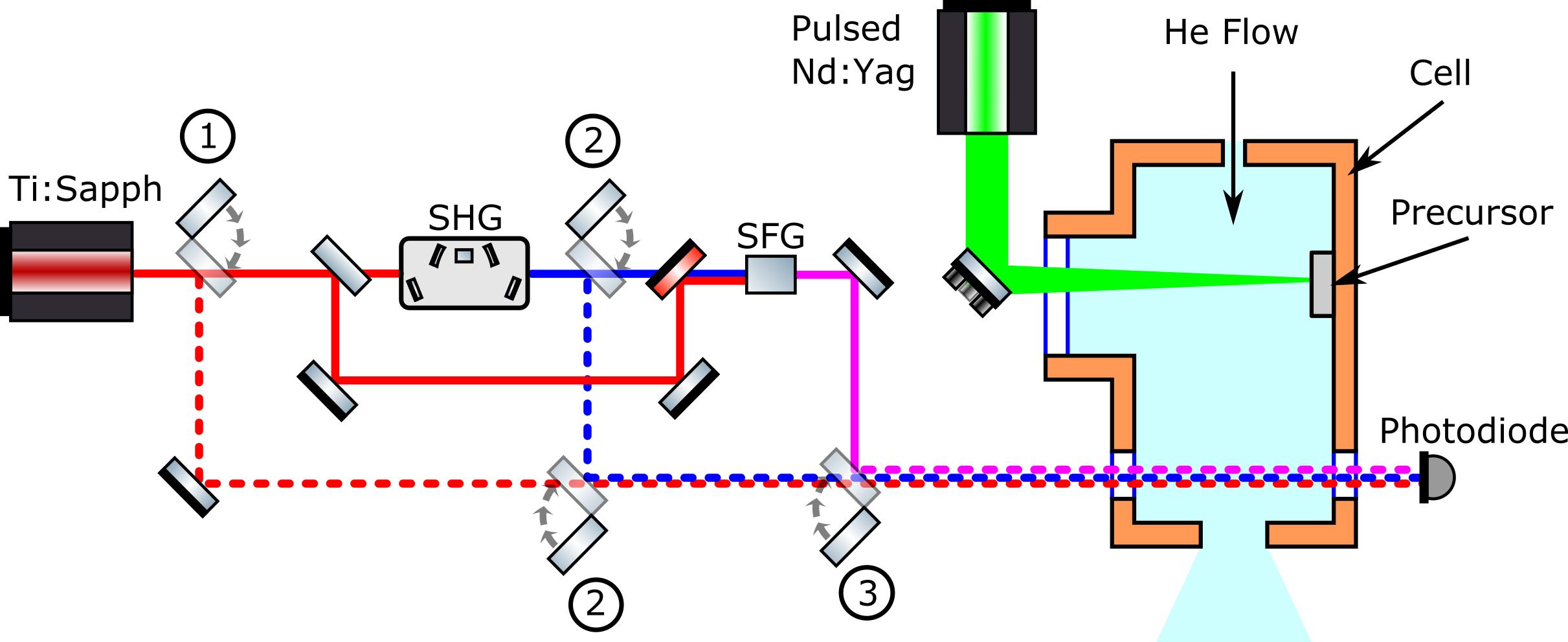}
        \caption{Schematic of the experimental apparatus. The flip mirrors (1,2 and 3) allow for the frequency range selection. Configuration (1) uses the Ti:Sapph output at 766\,nm (red) directly for measuring K. Configuration (2) uses the SHG output (bow-tie cavity with LBO crystal) at 395\,nm (blue) to measure Al and (3) uses the THG output from a BBO crystal at 261.5\,nm (violet) to measure AlCl absorption.}
    \label{fig:cell}
\end{figure}

The experimental apparatus is illustrated in Figure \ref{fig:cell} and details can be found in  reference.$\cite{Daniel2021}$  Gas phase AlCl is produced by laser ablation into a cryogenic buffer gas cell. The target is installed into the cell and cooled to 4.2\,K using a commercial cryogenic system (PT-420, Cryomech). We flow 4\,standard cubic centimeters per minute (sccm) of a purified helium buffer gas into the cell, resulting in a helium density of 1.75$\times 10^{15} atoms/cm^3$. This buffer gas is cooled to $\approx 4.2$\,K. The precursor target is ablated with 14\,mJ of 532\,nm pulsed laser light focused to an $\approx 80$\, micron diameter spot, with a repetition rate of 1\,Hz. The hot plume from the ablation collides with the buffer gas and is cooled to around 8.5\,K by the time it passes the spectroscopy window in the cell. The ablation laser spot is steered with an actuator mirror to average over inhomogeneities in the target surface and to avoid drilling holes into the target.  

For the absorption measurements of AlCl, we use continuous-wave UV laser light generated via second-harmonic generation (SHG) and sum-frequency generation (SFG) to frequency triple a Ti:Sapphire ring laser at 784\,nm. In the present configuration, we produce $\approx 30 \mu$W of 261.5\,nm laser light, which is sufficient for performing in-cell absorption measurements. For this study, we perform absorption spectroscopy on three different species for the targets, starting with AlCl, followed by Al, and then K. For each scan, the laser is tuned to the peak absorption of the corresponding species.  For K, we use the direct output of the Ti:Sapph and tune it to the $^2S_{1/2} \rightarrow\,
^2P_{3/2}$ transition at 767 nm ($\approx 391.0160$\,THz). 
For Al, we use the SHG of the Ti:Sapph and tune it to the $^2P_{1/2} \rightarrow\,
^2S_{1/2}$ transition at 395 nm ($\approx 759.9052$\,THz).
For AlCl, we tune the third-harmonic generation (THG) of the Ti:Sapph to the X$^1\Sigma^+, v = 0 \rightarrow A^1\Pi, v' = 0$ Q branch transition at 261.5 nm ($\approx 1146.33104$\,THz). By rastering the ablation beam over the target, an image of the absorption signal for each target and species is acquired (see Supplemental Information).
At each point on the sample, the transmitted signal $I$ is measured with an amplified photodiode (Thorlabs PDA25K2), and the optical density
%\begin{eqnarray}
$-\log_{10} \left( \frac{I_0}{I} \right),$
%\end{eqnarray}
where $I_0$ is the transmitted signal without any molecules present, is calculated. In order to generate an absorption value for the time-dependent absorption traces like those shown in Figure 2, the transmission signal is averaged over the time interval from 1-3 ms after the laser pulse.  For the signal traces in the following analysis, we average the absorption over at least 50 different laser spots on the target pellet.

\section{Results}
\subsection{Experimental Measurements of AlCl, Al, and K}
The ablation target consists of a pressed pellet of powdered Al and XCl$_n$ precursors.  For all cases, a robust, stable pellet was formed that could be cut and glued to the target plate.  Several Cl sources were tested in addition to AlCl$_3$, including NaCl, KCl, CaCl$_2$ and MgCl$_2$.  These chlorides are non-toxic and relatively stable.  MgCl$_2$ and CaCl$_2$ can absorb water to make a hydrate, but this process is very slow when the MgCl$_2$ or CaCl$_2$ is pelletized with Al, with less than a 0.03\% mass increase after 24 hours of exposure to ambient air. Of the other low molecular weight chlorides, LiCl is very hygroscopic and visibly changes appearance after less than 1 minute of air exposure, so it was not tested. BeCl$_2$ is toxic and was also omitted.  The properties of the alkali and alkaline earth chlorides are summarized in Table \ref{tab:alkaline}.  The variation in X-Cl ionic bond strengths provides a way to assess whether the initial dissociation of this bond is a limiting factor in AlCl production.
\begin{table}[h]
\small
\begin{tabular*}{0.48\textwidth}{@{\extracolsep{\fill}}lcccccc}\hline
& MW(g/mol) & $\rho$ (g/cm$^3$) & $\rho_{Cl}$ (mol/cm$^3$) & $D^0$ (kJ/mol)\\\hline
Al & 26.982	& 2.70 &	-	& -	\\
KCl & 74.551 &	1.988 &	0.0267 &	433.0 \\
NaCl & 58.443 &	2.17 &	0.0371 &	412.1 \\
CaCl$_2$ & 110.984 &	2.15 &	0.0387 &	409 (Ca-Cl)	\\
MgCl$_2$ & 95.211 &	2.325 &	0.0488 &	312 (Mg-Cl)	\\
AlCl$_3$ & 133.34 &	2.48 &	0.0558 &	502 (Al-Cl) \\\hline
\end{tabular*}
\caption{
\label{tab:alkaline}
Aluminum and XCl$_n$ sources in order of increasing chloride molar density. The molecular weights (MW), density ($\rho$), and bond dissociation energies ($D^0$) are also shown. $D^0$ values are shown for the diatomic bonds as denoted (X-Cl). \cite{crc2021}}
\end{table}

The apparatus for the production and measurement of the AlCl has been described in detail in a previous paper \cite{Daniel2021}. Briefly, a high-energy 14\,mJ, 5\,ns at 532\,nm laser pulse is focused on the target in a vacuum chamber.
The resulting ablation plume is swept through the absorption cell by a 4 K stream of He gas. The absorption at the $v=0,J=1 \rightarrow v'=0, J'=1$ transition located at 1146\,THz (261.5\,nm) gives rise to the most intense absorption and is monitored by an ultraviolet probe laser. 

\begin{figure}[h]
\centering
    \includegraphics[height=7cm]{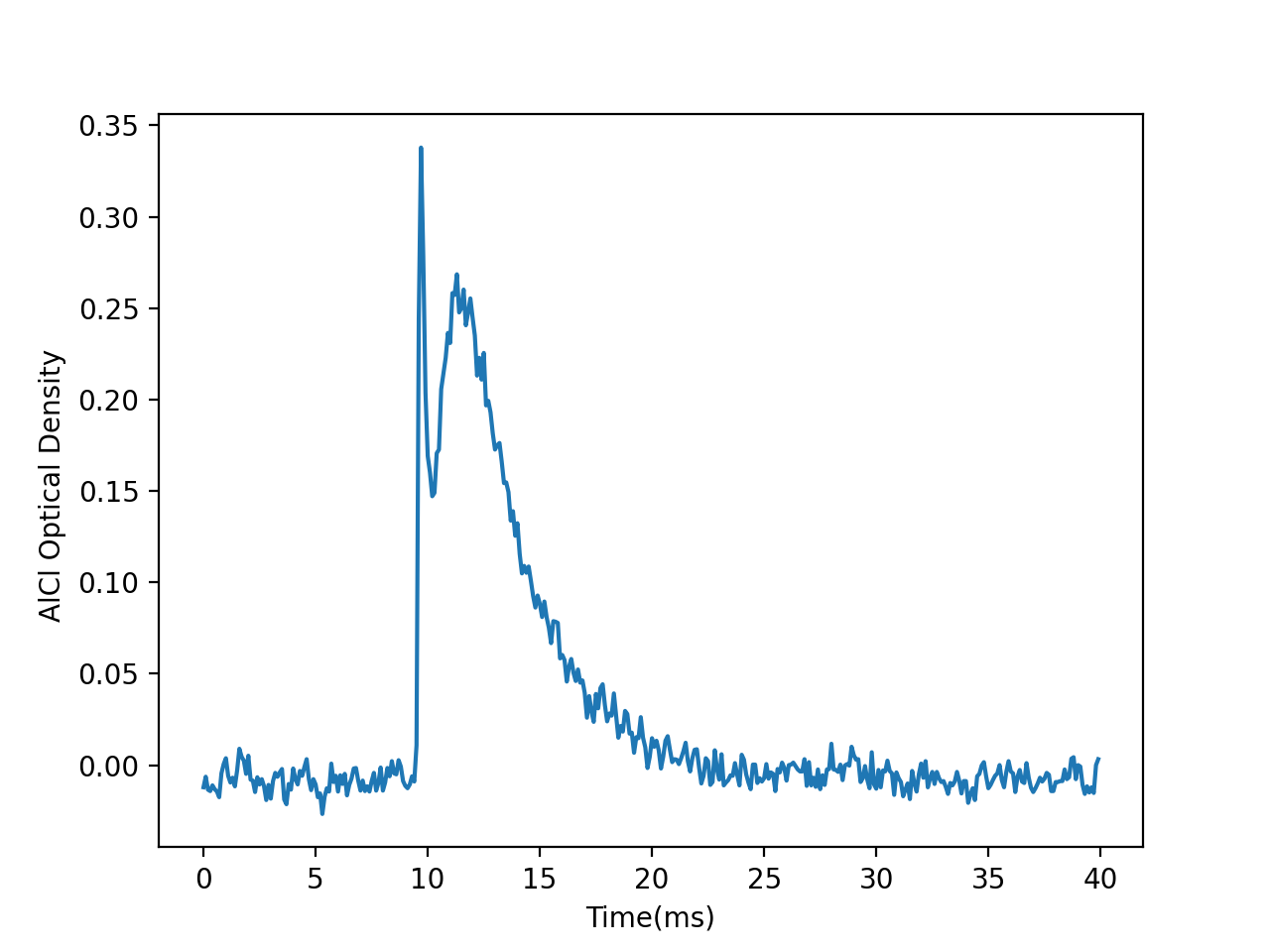}
    \caption{Sample absorption time trace at the peak of the Al$^{35}$Cl Q branch.
    We note that the initial spike at $t = 10$\,ms is an artifact from the ablation laser scatter on the absorption photodiode.}
    \label{fig:absorption_trace}
\end{figure}

\begin{figure}[h]
\centering
    \includegraphics[height=4cm]{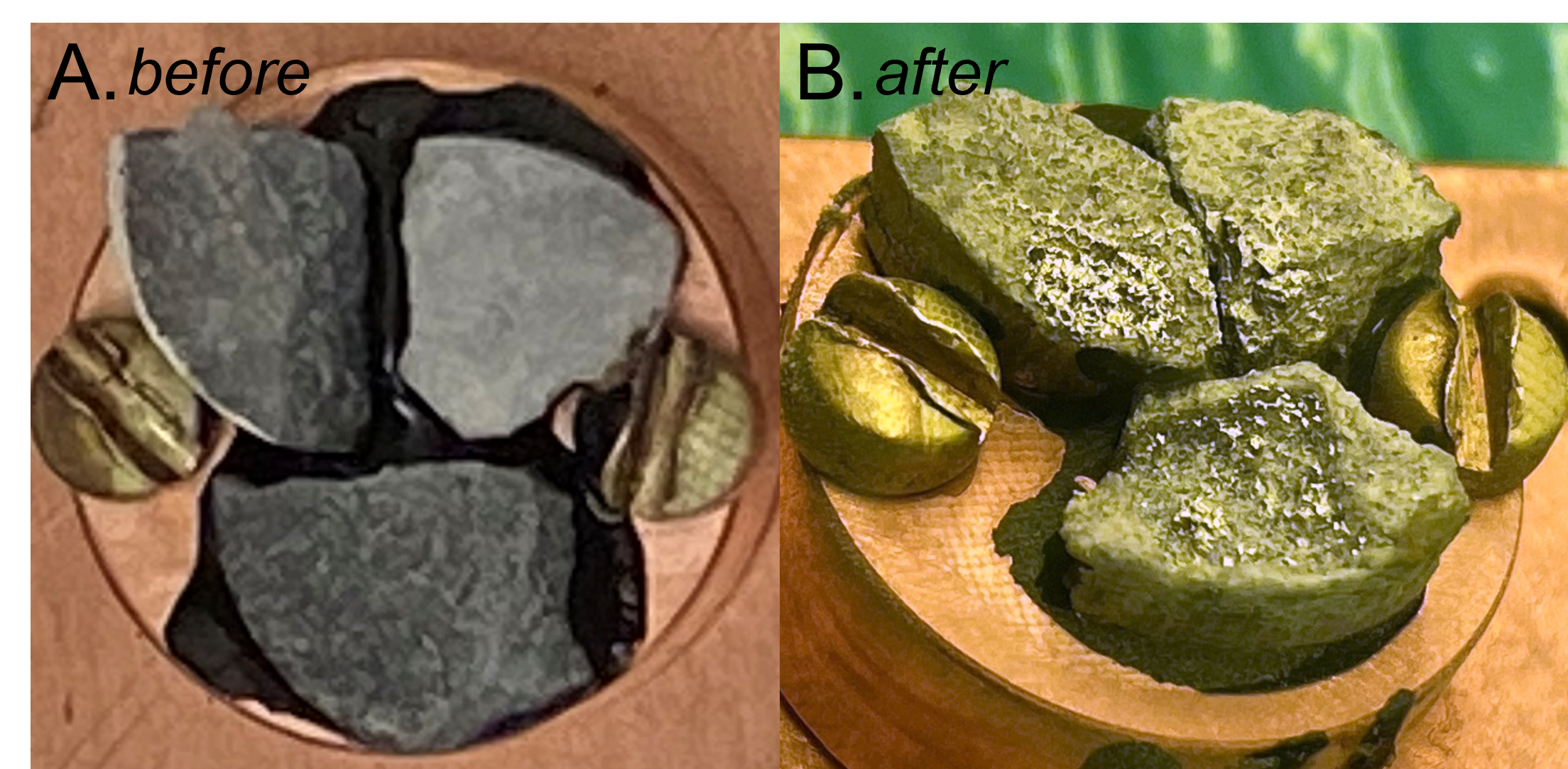}
    \caption{A) Typical sample targets of Al:KCl mixtures before ablation. Targets vary in shade of grey due to the amount of aluminum in Al:KCl molar ratio, consisting of pure KCl (top right), 1:3 (top left), and 1:10 (bottom). B) Typical target appearance after ablation to show the shiny aluminum coating on the sample after ablation. The Al:KCl ratios are 3:1 (top right), 1:1 (top left), and 1:3 (bottom).  
    \label{fig:targets}
     }
\end{figure}

The absorption is time-dependent, as shown in Figure \ref{fig:absorption_trace}, peaking shortly after the ablation laser shot at 10\,ms and then decaying away on a timescale of $\approx\,5$\,ms.  Figure \ref{fig:targets} shows images of a Al:KCl target before and after an absorption run. The grey color of the target in Figure \ref{fig:targets} can be attributed to the Al powder with its oxide coating.  The KCl does not absorb in the visible region and makes a transparent glassy solid when compressed.

After ablation, the target in Figure \ref{fig:targets}B shows the expected loss of mass, but this loss is not uniform across the pellet.  This corrugated landscape is probably due to local concentration inhomogeneities in the pellet, as well as morphology changes that occur due to fracture during the ablation process.  The most notable change is the appearance of a bright, reflective layer of Al metal in the ablated regions.  The presence of this coating suggests Al atoms are being efficiently ablated and recondensing onto the target surface as a metallic layer.  The effect of ablation on the KCl is less obvious, but it is also ablated as deduced from the large loss of material from the pellet.  From Figure 3, one potential concern is that debris from one target could contaminate a neighboring target.  However, this contamination would only affect the surface, while the majority of ablated material originates from the interior of the pellet.  Extra contributions by K atoms in the absorption signal of neighboring low KCl pellets was not detected, further indicating that cross-contamination is negligible.

\begin{figure*}
 \centering
 \includegraphics[height=4.5cm]{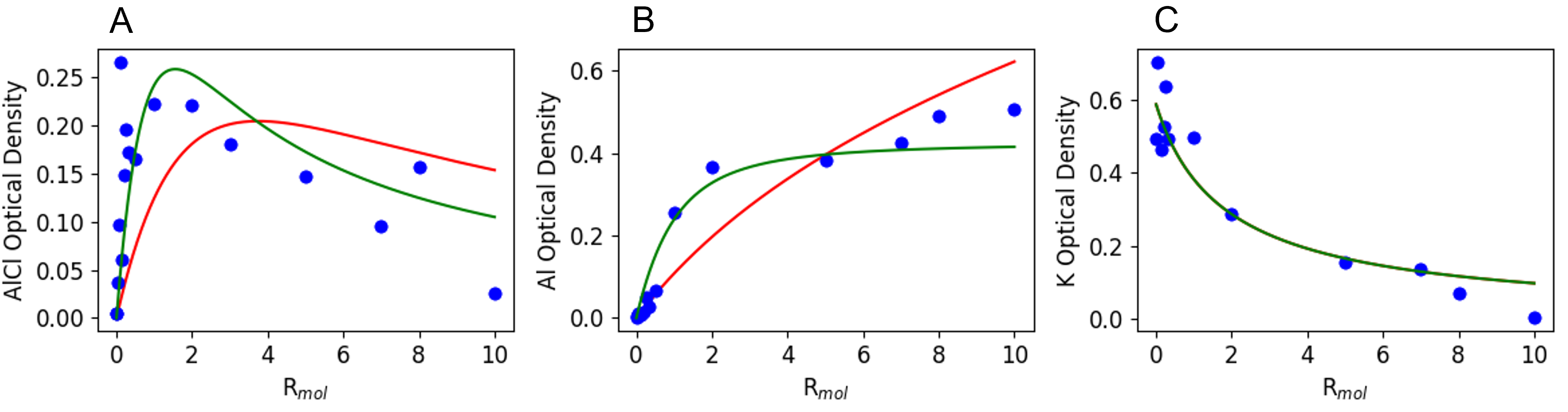}
\caption{
Model A (red line) and Model A' (green line) are overlaid with experimental data (blue dots) AlCl (A), Al (B), and K (C) showing a qualitatively better fit from Model A'. Model A and A' predict the same fit for K Optical Density in C. }
\label{model_data}
\end{figure*}

To determine how the AlCl production depends on the composition of the ablation target, the Al:KCl combination was used with varying amounts of KCl and Al powders.  The amounts of AlCl, K and Al produced from different mixing ratios were monitored in the absorption chamber. The experimental results are shown in Figure \ref{model_data} as a function of the Al:Cl molar ratio, $R_{mol}$.  Also shown in Figure \ref{model_data} are simulated data generated by two chemical models that are described in the next section. \\

\subsection{Modelling AlCl Formation}
Given the highly dynamic nature of the ablation process followed by rapid cooling in the He gas, it is doubtful that the system ever reaches chemical equilibrium.  The reaction
\begin{equation}
    Al + Cl \rightarrow AlCl
\end{equation}
is thermodynamically favored with $\Delta{G}=-75$ kJ/mol\cite{Numata1993}, but at low temperatures the reaction
\begin{equation}
    3AlCl \rightarrow 2Al + AlCl_3
\end{equation}
is even more favored.\cite{Russell1951,Kikuchi1964,Rao1966}  Extrapolating from previous high temperature results\cite{Kikuchi1964}, we can estimate K$_{eq}$=10$^{1000}$ for reaction (2) at 10 K, which implies that no AlCl should be present at this temperature if the system is at equilibrium.  But the disproportionation reaction (2) has an activation barrier and can be suppressed for temperatures below about 150 K, as shown in earlier work on cryogenic solids\cite{Tacke1989}. So the rapid cooling of the ablation plume should be able to quench reaction (2) and preserve the thermodynamically unstable AlCl.  Given that reaction (1) will be occurring under conditions of rapid cooling, AlCl formation will be determined by first-encounter collisions of Al and Cl atoms early in the process, followed by an extended period during which the excess collision energy present in the AlCl molecule is carried away by collisions with the He atoms.  Thus a reasonably high He gas density is expected to be necessary for preserving the newly formed AlCl\cite{Hutzler2012,Takahashi2021}.

To model the $R_{mol}$ data, we take as the starting point the assumption that the AlCl concentration [AlCl] is proportional to the product of the initial gas phase concentrations of Al and Cl atoms produced by the ablation pulse, [Al]$_0$ and [Cl]$_0$ respectively,
\begin{equation}
\label{reaction}
    [AlCl]=K[Al]_0[Cl]_0
\end{equation}
Note that the form of Equation \ref{reaction} resembles that expected for a system at equilibrium, where K=K$_{eq}$, the equilibrium constant.  But here Equation \ref{reaction} is justified by different physical considerations.  From a probabilistic standpoint, the collision probability is just proportional to the Al and Cl densities.  Alternatively, if we consider the bimolecular kinetic equation,
\begin{equation}
    \frac{\partial[AlCl]}{\partial t}=k[Al][Cl]
\end{equation}
where lower-case k is the rate constant, for small time intervals $dt$ we also obtain Equation \ref{reaction} with $K=kdt$.  These arguments justify the use of Equation \ref{reaction} as the starting point for our calculations.\\ 

\noindent
$\textbf{\emph{Model A: Free Atom Production}}$

We assume that the ablation laser vaporizes a volume $V_0$, which puts a limit on the amount of Al and Cl that can be vaporized because
\begin{equation}
\label{eq:density_limit}
    \frac{m_{Al}}{\rho_{Al}} + \frac{m_{XCl_n}}{\rho_{XCl_n}} = \frac{MW_{Al}}{\rho_{Al}} N_{Al} + \frac{MW_{XCl_n}}{\rho_{XCl_n}} N_{XCl_n} = V_0
\end{equation}
where $m_{Al}$ and $m_{XCl_n}$ are the masses of Al and ${XCl_n}$ in the ablation volume V$_0$. The molecular weights of the compounds are $MW_{Al}$ and ${MW_{XCl_n}}$; $\rho_{Al}$, $\rho_{XCl_n}$ are the solid-state densities of Al and XCl$_n$ and $N_{Al}$ and $N_{XCl_{n}}$ are the number of moles of atomic Al, and molecular XCl$_{n}$ in V$_0$.  

Equation \ref{eq:density_limit} places a constraint on the number of moles of Cl available to bond with a given number of moles of Al.  We define the molar ratio of Al:Cl atoms (R$_{mol}$) in the solid to be
\begin{equation}
\label{eq:rmol_A}
R_{mol} = \frac{1}{n} \frac{N_{Al}}{N_{XCl_n}} \quad .
\end{equation}
If we assume that the moles of atoms in the gas phase are directly proportional to the number of moles in the solid, i.e. $N_{Al}^\textrm{gas} = \alpha N_{Al}$ and $N_{Cl}^\textrm{gas} = \beta n N_{XCl_n}$ where $\alpha$ and $\beta$ are the ablation efficiencies and the $n$ factor takes into account that we get $n$ Cl atoms per molecule of $XCl_n$.   Equation \ref{reaction} then becomes
\begin{equation}
[AlCl] = K [Al]_0[Cl]_0 =
K \left(
\frac{N_{Al}^\textrm{gas}}{V_\textrm{gas}}
\right)
\left(
\frac{N_{Cl}^\textrm{gas}}{V_\textrm{gas}}
\right)
=
K \alpha \beta \left(
\frac{N_{Al}}{V_\textrm{gas}}
\right)
\left(
\frac{n N_{XCl_n}}{V_\textrm{gas}}
\right)
\end{equation}
Using Equation \ref{eq:density_limit} and Equation \ref{eq:rmol_A}, we obtain an expression for [AlCl], $N_\textrm{Al}^\textrm{gas}$ and $N_\textrm{Cl}^\textrm{gas}$ in terms of $R_{mol}$ and V$_0$ with only $\alpha$, $\beta$, and $\emph{K}$ as adjustable parameters:  
\begin{equation}
\label{eq:al}
N_\textrm{Al}^\textrm{gas} = n V_0 \left(\frac{\alpha}{\frac{MW_{XCl_n}}{\rho_{XCl_n}} + \frac{MW_{Al}}{\rho_{Al}} n R_{mol}}\right) R_{mol}
\end{equation}
\begin{equation}
\label{eq:cl}
N_\textrm{Cl}^\textrm{gas} = n V_0 \left(\frac{\beta}{\frac{MW_{XCl_n}}{\rho_{XCl_n}} + \frac{MW_{Al}}{\rho_{Al}} n R_{mol}}\right)\\
\end{equation}
\begin{equation}
\left[AlCl\right] =
\label{eq:alcl}
\gamma
\left(\frac{n}{\frac{MW_{XCl_n}}{\rho_{XCl_n}} + \frac{MW_{Al}}{\rho_{Al}} n R_{mol}}\right)^2 R_{mol}
\end{equation}
where $\gamma = K \alpha \beta
\left(\frac{V_0}{V_\textrm{gas}}\right)^2$.
Equation \ref{eq:alcl} gives the full dependence of the AlCl concentration on $R_{mol}$.  We can take the derivative of Equation \ref{eq:alcl} to find the $R_{mol}$ value that gives the maximum [AlCl], 
\begin{equation}
\label{eq:rmol_max_B}
R_{mol}^\textrm{max} = 
\frac{
\frac{MW_{XCl_n}}{\rho_{XCl_n}}
}{
n \frac{MW_{Al}}{\rho_{Al}}
}
\end{equation}
The maximum [AlCl] is found to be
\begin{equation}
[AlCl]_\textrm{max} =
\frac{\gamma}{4}
\frac{n \rho_{Al} \cdot \rho_{XCl_n}}{MW_{Al} \cdot MW_{XCl_n}}\quad.
\label{eq:alcl_max_B}
\end{equation}
Equation \ref{eq:rmol_max_B} shows that $R_{mol}^\textrm{max}$ depends only on the relative molar densities of Al and Cl atoms in the laser focal volume.  Equations 8-10 can be independently scaled to obtain agreement with the experimental data. Since the scaling factors may be different, this fitting does not permit the absolute determination of the efficiencies $\alpha$ and $\beta$.\\

\noindent
$\textbf{\emph{Model A': Al Recondensation}}$

The observation in Figure \ref{fig:targets} that metallic Al is plating out on the target suggests that the concentration of Al atoms above the target may be limited by recombination.  The relatively high boiling point of Al (2519$^{\circ}$C)\cite{crc2021} versus KCl (1407$^{\circ}$C)\cite{Kirshenbaum1962} suggests that the recondensation of Al into its liquid form will occur preferentially.  If Al atoms recondense before they can react with the Cl atoms, this will limit the Al concentration in the gas phase. We can take this possibility into account by limiting the amount of Al in the gas phase using the equation
\begin{eqnarray}
\label{eq:al_A'}
N_\textrm{Al}^\textrm{gas} = \frac{V_0\alpha}{\kappa} \left(
1 - e^{-\kappa N_{Al}/V_0}
\right)\quad,
\end{eqnarray}
$\kappa$ is a free parameter that limits the amount of Al that can react with Cl atoms. Equation \ref{eq:al_A'} assumes that the local density of Al in the solid, $\frac{N_{Al}}{V_0}$, promotes recondensation of the Al from the gas phase.  This expression reduces to N$_{Al}^\textrm{gas}=\alpha N_{Al}$ in the limit of small $\kappa$ or N$_{Al}$, but predicts that N$_{Al}^\textrm{gas}$ saturates at a value of $\frac{V_0\alpha}{\kappa}$ as N$_{Al}$ increases.  This expression leads to a new equation for the concentration of AlCl
\begin{eqnarray}
\label{eq:alcl_A'}
[AlCl]=
\gamma
\frac{n}{\kappa}
\frac{1}
{\frac{MW_{XCl_n}}{\rho_{XCl_n}}+
\frac{MW_{Al}}{\rho_{Al}} n R_{mol}}
\left[
1-\exp \left[
\frac{
-\kappa n R_{mol}}
{\frac{MW_{XCl_n}}{\rho_{XCl_n}}+
\frac{MW_{Al}}{\rho_{Al}} n R_{mol}}\right] 
\right]
\end{eqnarray}
To compare this result to Equation \ref{eq:rmol_max_B}, the maximum of Equation \ref{eq:alcl_A'} can only be determined numerically.
Physically, we expect the maximum of the [AlCl] curve to shift to lower R$_{mol}$ values, since increasing N$_{Al}$ only limits the available Cl atoms after saturation.  This is exactly what is observed in numerical calculations (see Supplemental Information).\\

\subsection{Comparison of Model Results with Data}

In Figure \ref{model_data}, we overlay the results of Models A and A' with the Al:KCl data.  The prefactors in Equation \ref{eq:alcl} and Equation \ref{eq:alcl_A'} are scaled to match the experimental data. Model A' does a better job of reproducing the data than Model A, which assumes no aluminum saturation after gas-phase collisions in the ablation plume.  Model A predicts a maximum signal for $R_{mol} = 3.75$, while Model A' predicts a maximum [AlCl] at $R_{mol} = 1.55$, with $\kappa = 53.2\, cm^3/mol$. Both Model A and Model A' predict the same trend in K signal, as shown in Figure \ref{model_data}C; while Model A' better predicts the trend in aluminum signal due to recondensation of aluminum after ablation, as shown in Figure \ref{model_data}B. Another advantage of Model A' is that it predicts a sharper drop off in the AlCl concentration for larger $R_{mol}$ values. The $\gamma$ prefactor required to match the [AlCl] data in Figure \ref{model_data}A is 2.8 times larger in Model A' than Model A. This corresponds to roughly three times lower [AlCl] in Model A' than Model A for a fixed $\gamma$ prefactor. Based on this trend, we estimate that the AlCl concentration is reduced by a factor of three due to recondensation of aluminum onto the surface of the target.  

\begin{figure}[h]
\centering
    \includegraphics[height=9cm]{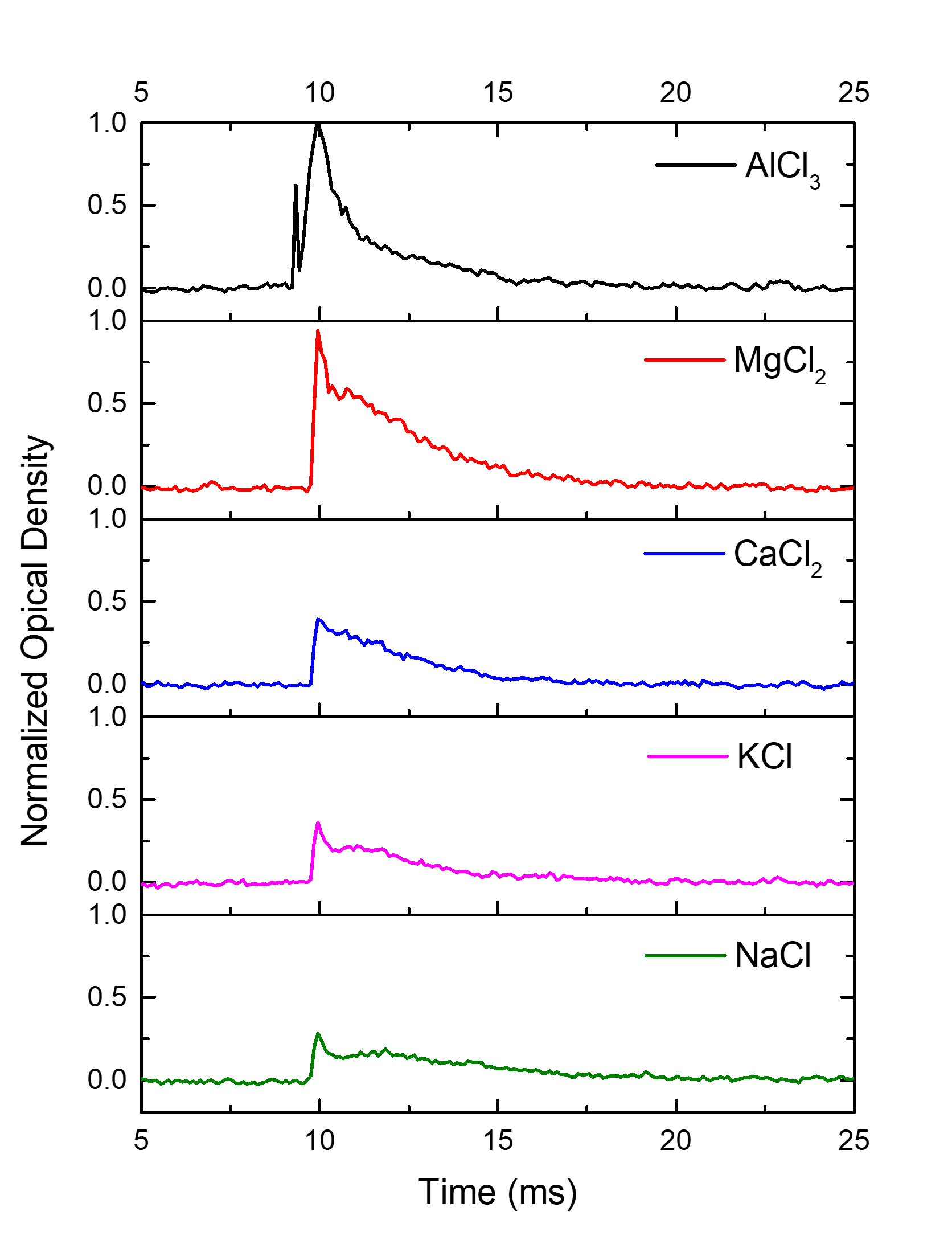}
    \caption{AlCl absorption after ablation of each chloride precursor at $R_{mol}=0.25$. Each spectra is normalized to the AlCl$_3$ optical density to show the decrease in signal in each chloride source from top to bottom. }
    \label{fig:absorption_comparison}
\end{figure}

\begin{figure}[h]
\centering
    \includegraphics[height=6.5cm]{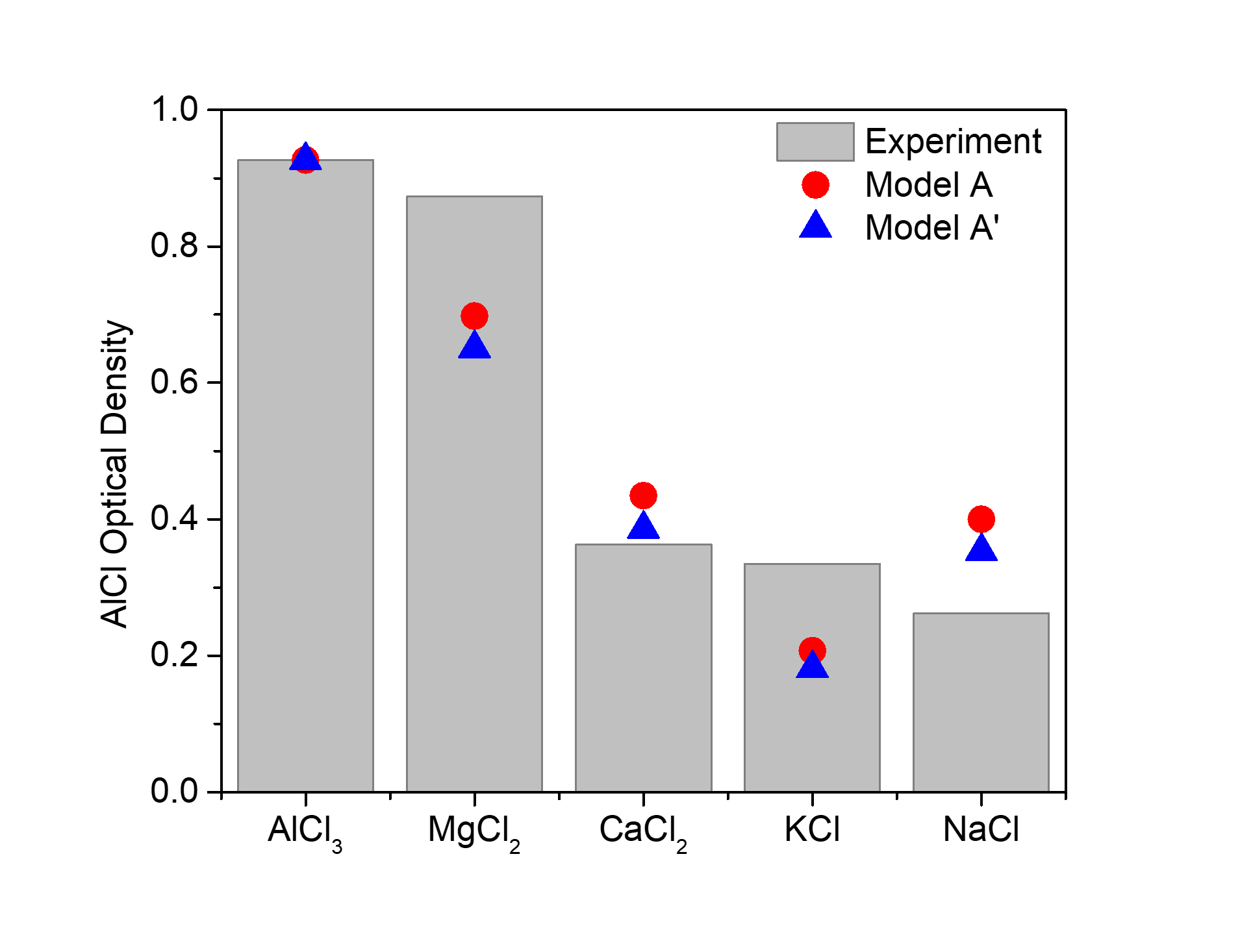}
    \caption{Theory is compared to the experimental optical density signal from each chloride source at $R_{mol}=0.25$. Models A and A' are normalized to the AlCl$_3$ optical density.}
    \label{fig:model_comparison}
\end{figure}

We next turn to the comparative study of different Cl atom sources.  Figure \ref{fig:absorption_comparison} compares the raw AlCl signals obtained for pure AlCl$_3$ and mixtures with NaCl, KCl, MgCl$_2$, and CaCl$_2$ with $R_{mol} = 0.25$. For this comparison, we used the best signal obtained after testing multiple AlCl$_3$ pellets, some of which produced no signal at all. The Al:XCl$_n$ mixtures, on the other hand, provided much more reproducible signal levels.  We observed only about a factor of two variation in the yield for these very different chemical mixtures.  This result was robust with respect to the method used to extract the average absorption from the time-dependent traces in Figure 5 (Supplemental Information).
Figure \ref{fig:model_comparison} provides a relative comparison of the experimental and calculated signals using models A and A'.  The XCl$_n$ species are ordered from low to high Cl molar density, as given in Table \ref{tab:alkaline}.   The general trends for all three models are very similar.  Both models predict a relatively weak dependence on the identity of X, as observed.

\begin{figure}[h]
\centering
    \includegraphics[height=11cm]{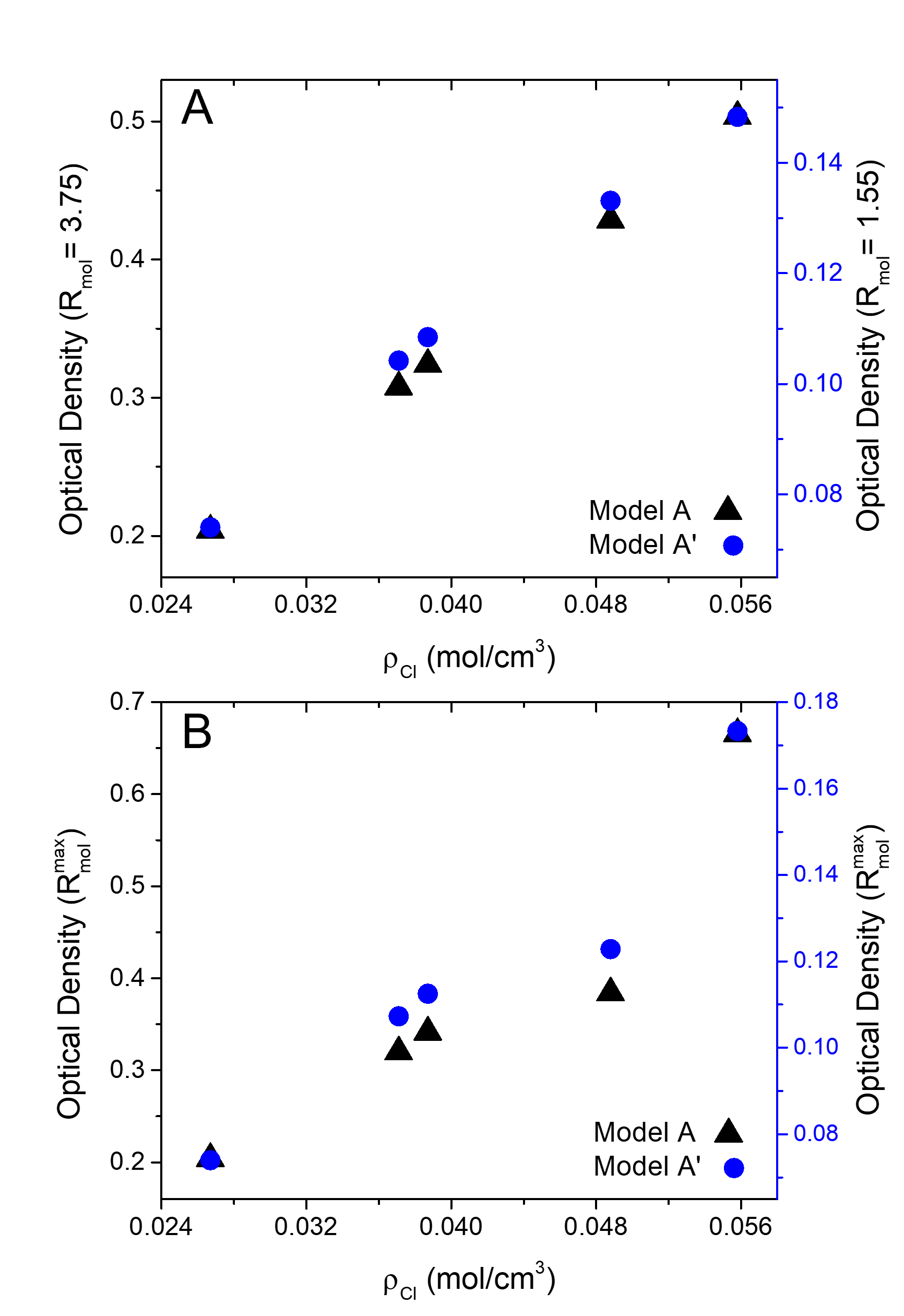}
    \caption{The chlorine molar density ($\rho_{Cl}$) for each XCl$_n$ source from Table \ref{tab:alkaline} is plotted against the optical density of that XCl$_n$ source from Model A and Model A'. A) The optical density for each XCl$_n$ source is taken at the $R_{mol}^\textrm{max}$ of KCl from each Model that is reported in Table S1 in the Supplemental Information. B) The optical density for each XCl$_n$ source is taken at its individual $R_{mol}^\textrm{max}$, as seen in the supplemental information. Both A and B show a linear dependence of AlCl concentration on $\rho_{Cl}$.}
    \label{fig:Cl_density}
\end{figure}

The AlCl yield is not correlated with variations in the X-Cl bond strength (Table \ref{tab:alkaline}), suggesting that the details of the chemical bonding in the XCl$_n$ precursor are not vital. Instead, the critical quantity for AlCl production appears to be the molar density of Cl atoms in the solid, $\rho_{Cl}$.  This can be seen most clearly when the calculated [AlCl] signal is plotted for each XCl$_n$ precursor at a fixed R$_{mol}$ value.  This plot, shown in Figure \ref{fig:Cl_density}A for models A and A', is linear and has the same slope for both models.  It should be noted that each XCl$_n$ source can have a different $R_{mol}^\textrm{max}$, so we also evaluated the maximum [AlCl] signal for each individual XCl$_n$ source.  The trend with $\rho_{Cl}$ is the same as in Figure \ref{fig:Cl_density}A, although not quite as linear as for the fixed $R_{mol}$ plot. In fact, this variation can be predicted from Equation \ref{eq:alcl_max_B} and the deviations from linearity are due to changes in the MW$_{XCl_n}$ denominator.  The most important conclusion is that variations between the different chlorides are almost entirely due to differences in Cl density in the solid. 

\section{Discussion}

Laser ablation of solid targets is an extremely complicated process, involving multiple processes like thermal melting, phase explosion, and ionization.\cite{Russo2013}  The laser fluences used in this experiment are well above the phase explosion\cite{Stafe2008,Gragossian2009,Zhang2017} and plasma thresholds \cite{Cabalin1998} for Al, suggesting that many different chemical species can be generated in the plume above the target.  Once the precursors are in the gas phase, modeling the detailed chemistry would require measuring densities, diffusion coefficients, and reaction rate constants for these species, a daunting task.  An important result of this paper is that relatively simple models can capture the main features of the AlCl production by considering only the dynamics of the Al and Cl atoms after they are placed into the gas phase.  The details of how they got there do not affect the model, as long as the ablation efficiencies are independent of $R_{mol}$.  This can be rationalized by assuming that the ablation conditions result in such high temperatures that essentially all solid-state bonding is lost, leading to free atoms whose reaction to form AlCl is entirely determined by the initial concentrations and collision rate. 

The fact that the first-encounter models do a good job of reproducing the [AlCl] curves is not surprising because the atoms are rapidly cooled as they react to form AlCl.  If the cooling process freezes out the initially formed AlCl and prevents subsequent high energy collisions that allow the molecules to react further, for example dissociating back into Al+Cl or by adding another Cl to make AlCl$_2$, then the mixture will not reach equilibrium.  At the measured temperature of 8.5\,K, the formation of AlCl$_3$ is overwhelmingly favored, so the fact that AlCl is observed at all is further evidence that the cryogenic buffer gas beam source does not allow the gas mixture to reach equilibrium.  In other words, AlCl production is kinetically controlled under these rapid cooling conditions. 
   
Having developed a physical model that does a reasonable job of quantitatively reproducing the AlCl production data, we are now in a position to draw some conclusions about the production of cold AlCl by laser ablation.  The first conclusion, that Cl atom density in the XCl$_n$ precursor is a critical parameter in determining AlCl yield, provides practical guidance for the choice of precursor.  In principle, AlCl$_3$ provides the highest yield, but as discussed above its toxicity and instability make it challenging to work with.  Given that Al:MgCl$_2$ provides a similar signal level, and since there is only a modest (roughly a factor of 2) variation between  the different chlorides we explored, it is not clear that the added inconvenience and hazard of AlCl$_3$ is worth the small signal gains.  On the other hand, it may be that molecules with additional bonded Cls could drive the yield even higher.  For example, hexachloroethane is a solid at room temperature, although it is much more difficult to handle than the salts used in this work.  Finally, we should emphasize that the physical basis of our model is very general and should be applicable to other halides.  We suspect that identifying precursors with high solid-state halide densities will be an important consideration for the production of metal halide diatomics using laser ablation from mixtures.

In addition to the choice of Cl source, other experimental parameters can be tuned to improve AlCl production.  One obvious step would be to increase the spot size and thus V$_0$, while keeping the heat load on the cryogenic system manageable. Since [AlCl] will scale as V$_0^2$ (Equation \ref{eq:alcl}), this provides a straightforward path to more signal per shot, albeit at the expense of using up the pellet more quickly and increasing the heat load on the cryogenic cell. Similarly, increasing the ablation efficiencies $\alpha$ and $\beta$ (which were not directly measured in this paper) would also increase the production of AlCl.  A comparison of the measured Al density ($1.2x10^9 atoms/cm^3$) and the maximum Al density calculated based on the estimated ablation volume ($1.1x10^{13} atoms/cm^3$, see Supplemental Information for details) allows a rough estimate of $\alpha$ to be on the order of $10^{-4}$.  Modifying the ablation conditions might enhance efficiencies.  For example, most researchers use ultraviolet excimer laser sources to ablate alkali halides\cite{Haglund1992,ThomasDickinson1994,Nishikawa2000,FA2008}, although infrared lasers have also been used\cite{Redeker2017}, so it is possible that shifting to shorter wavelengths would produce more Cl atoms.  Alternatively, if we assume that the Al component is responsible for most of the laser absorption, then smaller Al and XCl$_n$ particles in combination with more uniform mixing might also improve heat transfer to the XCl$_n$ and accelerate its solid-to-gas transition.  We could also consider ways to avoid the Al saturation behavior that is described by Model A'.  Changes in the rate of cooling gas flow or surface geometry might inhibit Al atom recombination, although this is somewhat speculative.

 Finally, it is important to point out that ablation of an Al:XCl$_n$ mixture will always be limited by the constraint of Equation \ref{eq:density_limit}.  Because all the Al and Cl atoms must be packed into a fixed volume V$_0$, increasing N$_{\rm Al}$ requires decreasing N$_{\rm Cl}$ and vice versa.  To decouple these quantities requires separate Al and Cl sources. Recent work has shown that using a gas source for the halide, for example SF$_6$, can successfully generate metal halides like AlF.\cite{Aggarwal2021}  To produce a source of Cl atoms, Cl$_2$, HCl or possibly methanochlorides like CCl$_4$, CHCl$_3$ and CH$_2$Cl$_2$  would be reasonable candidates.
 There is a previous report of AlCl being produced by ablation of an Al rod exposed to Cl$_2$ gas \cite{Hensel1993}, but its characteristics were not described in detail.  These potential Cl sources are corrosive and/or toxic, so the introduction of these gases into a vacuum chamber would add experimental challenges.  But they are also chemically stable and easy to put in the gas phase, so they could result in much higher AlCl production if all the solid Al could be vaporized and reacted.

\section{Conclusions}
This work has demonstrated that pulsed laser ablation of Al:XCl$_n$ mixtures provides a robust and reliable source of cold AlCl molecules.  Stable, non-toxic precursors can be used instead of AlCl$_3$, the most commonly used precursor in previous studies.  The reason that the alkali halide mixtures are relatively insensitive to the chemical nature of the precursor is that high intensity laser ablation provides enough excess energy to efficiently dissociate the Cl salt into its atomic constituents.  A simple model that assumes AlCl formation is mainly determined by the initial Al+Cl $\rightarrow$ AlCl encounters can quantitatively capture trends in the AlCl production as a function of precursor composition and Al:XCl$_n$ mixing ratio.  The most important attribute of the solid XCl$_n$ source is a high Cl atom density, a conclusion that may be generalizable to the production of other heteronuclear diatomics as well.  More powerful lasers, improved ablation of the Al component, and decoupling the Al and Cl sources are all promising future directions for producing a large numbers of cold AlCl molecules. The work in this paper represents a first step in understanding the chemical mechanisms of laser ablation sources for producing AlCl and will hopefully provide guidance for their future development and optimization.

\section*{Author Contributions}
Taylor N. Lewis and Chen Wang contributed equally to this work.

\section*{Conflicts of interest}
There are no conflicts to declare.

\section*{Acknowledgements}
We acknowledge funding from the National Science Foundation (NSF) RAISE-TAQS grant 1839153.

%%%END OF MAIN TEXT%%%


\balance


\renewcommand\refname{References}

%%%REFERENCES%%%
\bibliography{references}
\bibliographystyle{rsc} %the RSC's .bst file

\end{document}